\documentclass[final,3p,times,twocolumn,sort&compress]{elsarticle}
\usepackage{amssymb}
\usepackage{amsmath}
\usepackage{xcolor}
\usepackage{mathtools}
\usepackage[hidelinks]{hyperref}
\usepackage{xr}

\makeatletter
\newcommand*{\addFileDependency}[1]{% argument=file name and extension
\typeout{(#1)}% latexmk will find this if $recorder=0
\@addtofilelist{#1}
\IfFileExists{#1}{}{\typeout{No file #1.}}
}\makeatother

\newcommand*{\myexternaldocument}[1]{%
\externaldocument{#1}%
\addFileDependency{#1.tex}%
\addFileDependency{#1.aux}%
}

\myexternaldocument{supp_JoPSA_3cells_2}

\journal{Journal of Power Sources Advances}

\begin{document}

\begin{frontmatter}

\title{Effects of High-Temperature Preconditioning on Li Plating During Low-Temperature Cycling of the COTS LG Chem HG2 Cell}

\author[a,b]{Lukas~Lehnert}
\author[c]{Keith~B.~Chin}
\author[c]{Frederick~C.~Krause}
\author[c]{John~P.~Ruiz}
\author[a,b]{Simon~Hein}
\author[c]{John~Bescup}
\author[c]{Gil~Garteiz} 
\author[a,b,d]{Arnulf~Latz}
\author[c]{Erik~J.~Brandon}
\author[a,b,d]{Birger~Horstmann\corref{cor1}}

\cortext[cor1]{Corresponding author: birger.horstmann@dlr.de}

\affiliation[a]{organization={German Aerospace Center},%Department and Organization
            addressline={Wilhelm-Runge-Straße~10}, 
            city={Ulm},
            postcode={89081}, 
            %state={},
            country={Germany}}
            
\affiliation[b]{organization={Helmholtz Institute Ulm for Electrochemical Energy Storage~(HIU)},%Department and Organization
            addressline={Helmholtzstraße~11}, 
            city={Ulm},
            postcode={89081}, 
            %state={},
            country={Germany}}
            
\affiliation[c]{organization={Jet Propulsion Laboratory—California Institute of Technology},%Department and Organization
            addressline={4800 Oak Grove}, 
            city={Pasadena},
            postcode={91109}, 
            state={CA},
            country={USA}}
            
\affiliation[d]{organization={University of Ulm},%Department and Organization
            addressline={Albert-Einstein-Allee~47}, 
            city={Ulm},
            postcode={89081}, 
            %state={},
            country={Germany}}

%% Abstract
\begin{abstract}
Commercial off-the-shelf~(COTS) Li ion cells offer an attractive and rechargeable power source for spacecraft applications. However, since these cells are typically designed for charging above 0\,\textdegree C, challenging sub-zero temperatures could inflict irreversible degradation and safety hazards during in-flight operation. To address these challenges, this work explores strategies affecting their low-temperature performance. Specifically, we investigate the Li plating behavior of the COTS~LG~Chem~18650~HG2 cell during galvanostatic cycling at $-20\,$\textdegree C and its dependence on prior high-temperature preconditioning protocols, involving either two-week galvanostatic cycling or storage at 30\,\textdegree C, as well as a reference case without preconditioning, i.e., direct cycling at $-20\,$\textdegree C. Experimental cycling tests reveal distinct differences in Li metal deposition between cells subjected to these conditions. To identify the origin of these discrepancies, we combine computed tomography of the electrode microstructure with three-dimensional, spatially resolved cell modeling and analyze the simulation results.
\end{abstract}



%% Keywords

\end{frontmatter}



\section{Introduction}
Powering spacecraft is an ambitious task. Harsh and extreme environments demand challenging requirements from deployed energy and battery systems. Despite these demands, Li ion batteries have repeatedly demonstrated their space applicability~\cite{Jones2022, Krause2021, Krause2020}. Many space missions -- including the Mars Exploration Rovers Spirit and Opportunity, the Juno mission to Jupiter, and the InSight lander on Mars -- employed custom Li ion cells specifically designed to meet the desired battery characteristics. In recent years, however, interest has transitioned towards the use of commercial-off-the-shelf~(COTS) cells in the 18650 format, which are widely used in Low-Earth Orbit~(LEO) satellites and have been successfully deployed in missions such as the Ingenuity Helicopter on Mars and the MarCO CubeSats~\cite{Jones2022, Krause2021}.\\
The COTS cells, while not specifically designed for space environments, are preferred wherever possible before transitioning to custom solutions due to several advantages, even in low-temperature applications. Although commercial cells can be modified with custom electrolytes, manufacturers typically do not incorporate them into their peak-performing cells with the highest specific energy. Therefore, potential improvements in low-temperature performance may come at the cost of reduced overall capacity and energy due to lower performing cell chemistries. Furthermore, custom cells are generally not produced on the highest volume manufacturing lines, and therefore may lack the performance optimization gained from mass production, potentially leading to more variability in the cell characteristics. In contrast, COTS cells benefit from automated, high-volume manufacturing, ensuring high consistency and uniformity. This allows for replacing elaborate cell balancing electronics by a less expensive screening and matching process for the individual cells of a battery pack~\cite{Krause2020, Krause2021, pearson2005small, Jones2022}. Additionally, COTS 18650 cells offer safety features, high specific energies and energy densities, modularity, redundancy, and lower cost~\cite{Krause2020, Krause2021, Jones2022}.\\
However, COTS cells are often designed for charging between $0\leq T\leq 60\,$\textdegree C and thus, not explicitly intended for low-temperature operation -- a key requirement for many space missions, especially for missions exploring deep space~\cite{Chin2018, Gupta2020, Jones2022}. Low temperatures slow the intrinsic kinetics of the cell, which results in reduced capacity retention and rate capability, and increased risk of Li plating on the anode surface~\cite{Gupta2020, Chin2018, Krause2021}. Plated metallic Li not only aggravates the cell durability and cycle life but poses safety hazards and should be avoided~\cite{Waldmann2018, Hein2016, Gupta2020}.\\ 
This study seeks to expand the performance of COTS cells during reduced temperature cycling through simulations and experimental investigation of the LG Chem HG2 18650 cell, following different preconditioning protocols. This cell has already shown promising results, especially in maintaining high specific energy densities at elevated discharge rates down to $-20$\,\textdegree C~\cite{Krause2021}.\\ 
Herein, we examine the Li plating-related behavior at $-20$\,\textdegree C. Experimental cycling tests suggest that prior high-temperature preconditioning systematically affects the electrochemical response during subsequent low-temperature cycling. Cells that were simply stored at 30\,\textdegree C or cycled directly at $-20$\,\textdegree C without prior preconditioning exhibit reversible stripping signatures during subsequent cycling at $-20$\,\textdegree C. In contrast, cells that were subjected to preconditioning cycling at 30\,\textdegree C show consistently improved cycling metrics within the first 60 cycles, along with suppressed stripping signatures under the investigated conditions.\\
In order to deeply understand the underlying processes, we combine a three-dimensional, spatially resolved cell model with computed tomography~(CT) data, imaging the cell electrodes. The model simulates cycling and incorporates the formation of the solid-electrolyte interphase~(SEI) on the anode, inducing degradation and capacity fade. Calculating the cell state variables allows for tracking the overpotential, which drives the Li metal deposition reaction at the anode. With this, we provide spatially and temporally resolved predictions, if Li plating is thermodynamically promoted and thus, likely to occur, without explicitly modeling the Li plating process. \\
The following section~\ref{experiments} outlines the experimental procedures for the LG~HG2~cell, including different preconditioning protocols -- high-temperature cycling, high-temperature storage, and no preconditioning -- followed by low-temperature cycling tests, and presents the observed evidence of Li metal deposition. Section~\ref{simulations} introduces our cell model and describes the simulations, which are designed to mirror the experiments. To explore the origin of the Li plating behavior observed in the experiments, we analyze our simulation results and investigate the time evolution of spatially resolved key state variables. Section~\ref{discussion} compares the experimental and numerical results and discusses possible causes of observed deviations.

\section{Experiments}
\label{experiments}
\subsection{Methods}
\label{Methods_Experiments}
We experimentally examine the effects of different high-temperature preconditioning protocols on subsequent low-temperature cycling using the COTS LG HG2 cell. The first protocol~(cell A$_\mathrm{exp}$) consists of a two-week cycling procedure at 30\,\textdegree C, which comprises $n=2$ consecutive sets of $N=15$ cycles, separated by a multi-hour rest period. The second protocol~(cell B$_\mathrm{exp}$) involves simply storing the cell at 30\,\textdegree C for the same total duration. The third protocol~(cell C$_\mathrm{exp}$) omits any high-temperature preconditioning, and the cell is cycled directly at $-20$\,\textdegree C. While cell A$_\mathrm{exp}$ begins its preconditioning cycling with a cell potential of $U=3.9\,$V, cell B$_\mathrm{exp}$ and cell C$_\mathrm{exp}$ are both initialized at $U=3.5\,$V.\\ 
After the high-temperature preconditioning, all cells are subjected to low-temperature cycling at $-20\,$\textdegree C, consisting of several consecutive sets of $N=15$--25 cycles, again separated by multi-hour rest periods. Note that at both temperatures, each cycling set includes an extra discharge at the beginning and an extra charge at the end.\\
For each protocol, two cells were tested to ensure reproducibility. The cells were stored and cycled in a controlled temperature environment~(ESPEC North America, Inc., Platinous Chambers), using a custom-designed High Precision Coulometry instrument featuring a high precision digital multimeter~(Keithley 2000 Digital Multimeter) and a programmable current source~(Keithley 2400 SourceMeter, 200 V, 1 A, 20 W, Series 2400) operated using LabView. Both the high- and low-temperature cycling protocols followed the same procedure: galvanostatic cycling between 3.0\,V and 4.2\,V, with a 300\,s rest period separating the charge and discharge processes. The current amplitude was set to $I=0.6\,$A, which corresponds to a rate of C/5 according to the name plate cell capacity of 3\,Ah. \\ 
In addition, we resolve the microstructure of the cell electrodes using CT at a spatial resolution of 320\,nm, assigning each voxel to the corresponding electrode material based on the measured grayscale values~(see SI Section~S3.1). These reconstructed structures are later incorporated into our three-dimensional simulations, which are used to numerically predict Li plating within the investigated cells~(see Section~\ref{simulations}).
\begin{figure*}[p!]
\begin{center}
\includegraphics[width=\linewidth]{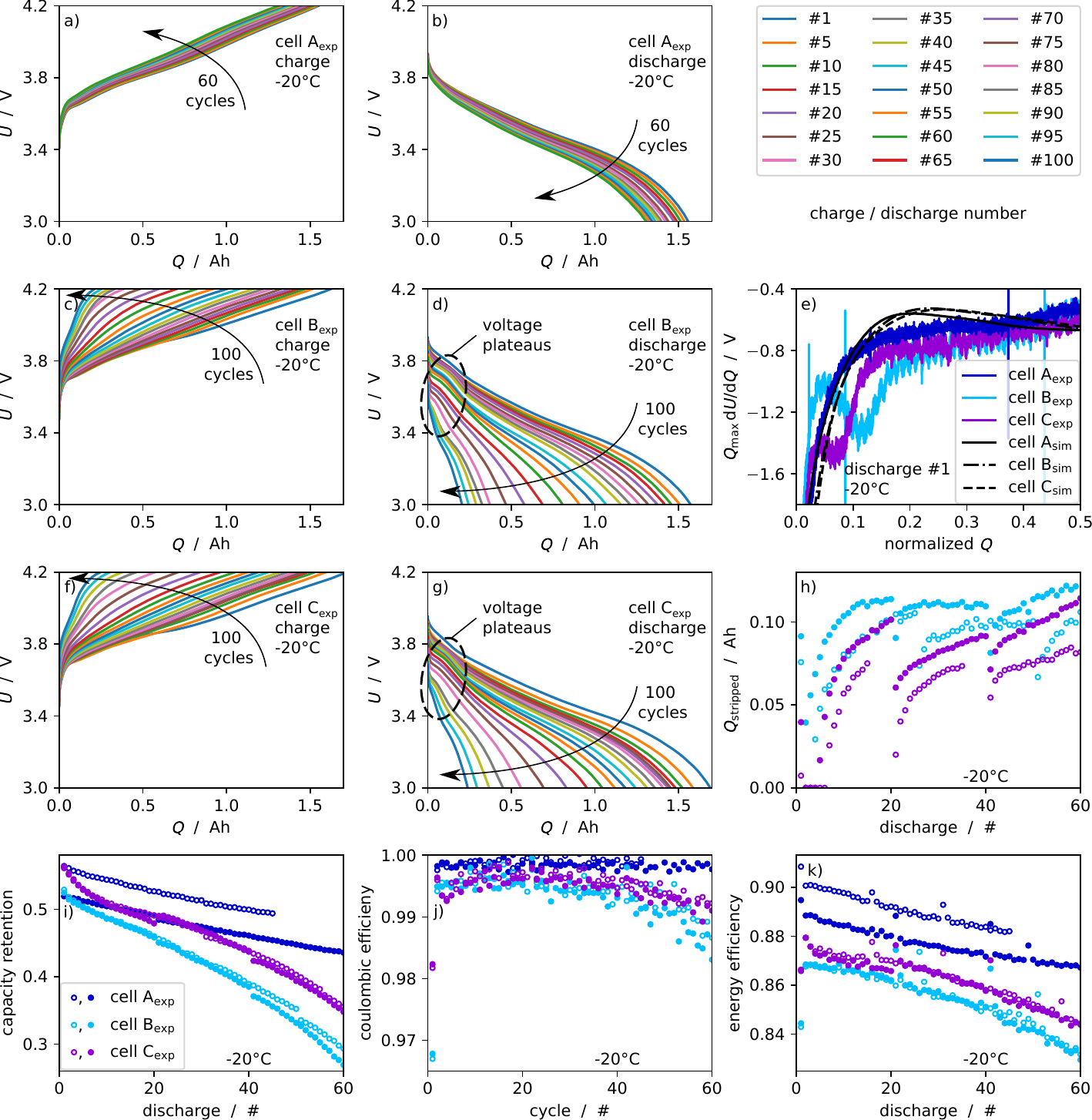}
\caption{a)--d), f), g) Exemplary experimental charge and discharge curves, e) exemplary differential voltage analysis as a function of capacity, h) stripped Li capacity $Q_\mathrm{stripped}$ during discharge, i) capacity retention, j) coulombic efficiency~(CE), and k) energy efficiency of the low-temperature cycling for cells A$_\mathrm{exp}$, B$_\mathrm{exp}$, and C$_\mathrm{exp}$. Panel j) is restricted to CE values $\leq1.0$ to highlight the small but systematic differences between the cells. Consequently, outliers occurring at the beginning of cycling sets are omitted. A corresponding CE plot including values above 1.0 is provided in the supplementary information~(see SI Section~S1.2). All quantities shown in panels h)--k) are evaluated over the common cycle window of the data sets. The cells corresponding to A$_\mathrm{exp}$ were measured up to 45 and 60 cycles, respectively, whereas the cells corresponding to B$_\mathrm{exp}$ and C$_\mathrm{exp}$ were measured up to 100 cycles. For comparability, only the overlapping cycle range is shown. Full-cycle data for B$_\mathrm{exp}$ and C$_\mathrm{exp}$ are provided in the supplementary information~(see SI Section~S1.2).} 
\label{pic_CE}
\end{center}
\end{figure*}
\subsection{Results}
\label{exp_battery_cycling}
Figure~\ref{pic_CE} summarizes the experimental cycling results. This includes exemplary charge and discharge processes, a corresponding differential voltage~(DV) analysis normalized by the measured cell capacity $Q_\mathrm{max}$~(i.e., $Q_\mathrm{max}$\,d$U$/d$Q$ plotted against normalized capacity $Q$), the resulting amount of stripped Li $Q_\mathrm{stripped}$ during discharge, as well as the capacity retention, coulombic efficiency~(CE), and energy efficiency of the low-temperature cycling for the cells A$_\mathrm{exp}$, B$_\mathrm{exp}$, and C$_\mathrm{exp}$. As evident in Figure~\ref{pic_CE}, the cells show deviating behavior.\\
Both cells B$_\mathrm{exp}$ and C$_\mathrm{exp}$ exhibit indications of voltage plateaus at the beginning of their discharges. These plateaus may be a signature of Li stripping and, consequently, suggest prior Li plating~\cite{Janakiraman2020, Smart2011, Bugga2010}, as they can arise from a phase equilibrium between plated Li and the first deintercalation stage of graphite~\cite{Petzl2014}. However, the discharge voltage can also be affected by thermal effects such as self-heating, short-lived local variations in electrode SOC at the very beginning of discharge, and concentration gradients within active particles, all of which complicate the direct interpretation of voltage plateaus~\cite{Campbell2019}.\\
The DV analysis provides a more robust approach to disentangle these contributions and identify Li stripping. While the temperature- and concentration-driven effects typically relax over a larger portion of the discharge, pronounced minima in the DV profile correspond to phase transitions in electrode materials~\cite{Campbell2019}. When comparing the experimental DV curves to numerical predictions that do not include Li plating~(see Section~\ref{simulations}), the DV profiles of cells B$_\mathrm{exp}$ and C$_\mathrm{exp}$ show minima at the onset of discharge that are absent in the simulated profiles~(see Fig.~\ref{pic_CE}e). This discrepancy indicates that these features likely arise from Li stripping associated with prior Li plating.\\
To quantify the amount of stripped Li $Q_\mathrm{stripped}$, we analyzed the DV profiles of the cells in which local minima indicative of Li stripping were visually observed. Around these minima, the profiles were fitted using a 10th-order polynomial, and the local roots of $Q_\mathrm{max}$\,d$^2U$/d$Q^2$ were determined~(see SI~Section~S1.1). These roots define the start ($Q_\mathrm{s}$) and end ($Q_\mathrm{e}$) of the Li stripping processes, allowing $Q_\mathrm{stripped}$ to be calculated as $Q_\mathrm{e}-Q_\mathrm{s}$~\cite{Campbell2019}. The resulting values of $Q_\mathrm{stripped}$ are shown in Figure~\ref{pic_CE}h.\\
As evident in Figure~\ref{pic_CE}h, cells B$_\mathrm{exp}$ and C$_\mathrm{exp}$ exhibit a non-monotonic evolution of stripped Li during the first low-temperature cycling set. A relatively high amount of $Q_\mathrm{stripped}$ is observed in the first discharge, followed by a decrease in the subsequent cycle. A gradual increase is then observed with continued cycling -- starting between the second and seventh discharge, depending on the cell -- which eventually exceeds the initial value. This behavior suggests significant Li plating during the first charge, reduced plating in the following cycle, and increasing plating with further cycling. However, it should be noted that $Q_\mathrm{stripped}$ provides only an approximate lower-bound estimate of the total amount of previously plated Li, as it accounts solely for the reversibly stripped fraction. Plated Li that has chemically intercalated into the anode, lost electrical contact, or reacted to form SEI species is not captured~\cite{Campbell2019}.\\
Similar trends are observed in the subsequent cycling sets, where $Q_\mathrm{stripped}$ again increases gradually with cycling, but without the pronounced initial peak observed in the first set. This suggests that metallic Li may be lost during the rest periods between sets, consistent with the formation of irreversible Li deposits or side reactions.\\ 
Additionally, cells B$_\mathrm{exp}$ and C$_\mathrm{exp}$ show lower capacity retention, reduced CEs, and decreased energy efficiencies compared to cell A$_\mathrm{exp}$. These observations suggest increased internal resistances and the occurrence of parasitic side reactions, irreversibly consuming Li$^+$ ions and hence reducing the cell capacity~\cite{Janakiraman2020}.\\
In contrast, cell A$_\mathrm{exp}$ exhibits no voltage plateaus in the discharge curves, no corresponding $Q$\,d$U$/d$Q$ peaks, and shows higher capacity retention, elevated CEs, and increased energy efficiencies, none of which provide direct indication of Li plating. We interpret these observations as evidence of little or no Li plating, suggesting that the high-temperature preconditioning protocol may influence Li plating behavior during low-temperature cycling. This effect may stem from differences in the induced cell~SOC at the beginning of the low-temperature cycles, as will be demonstrated later in the simulation results~(see Section~\ref{simulations}), or from other factors, as discussed in Section~\ref{discussion}. It should be noted, however, that the absence of observable voltage plateaus or DV features does not guarantee the absence of Li plating~\cite{Campbell2019}, and this interpretation should therefore be treated with caution.

\begin{figure*}[!t]
\begin{center}
\includegraphics[width=0.9\linewidth]{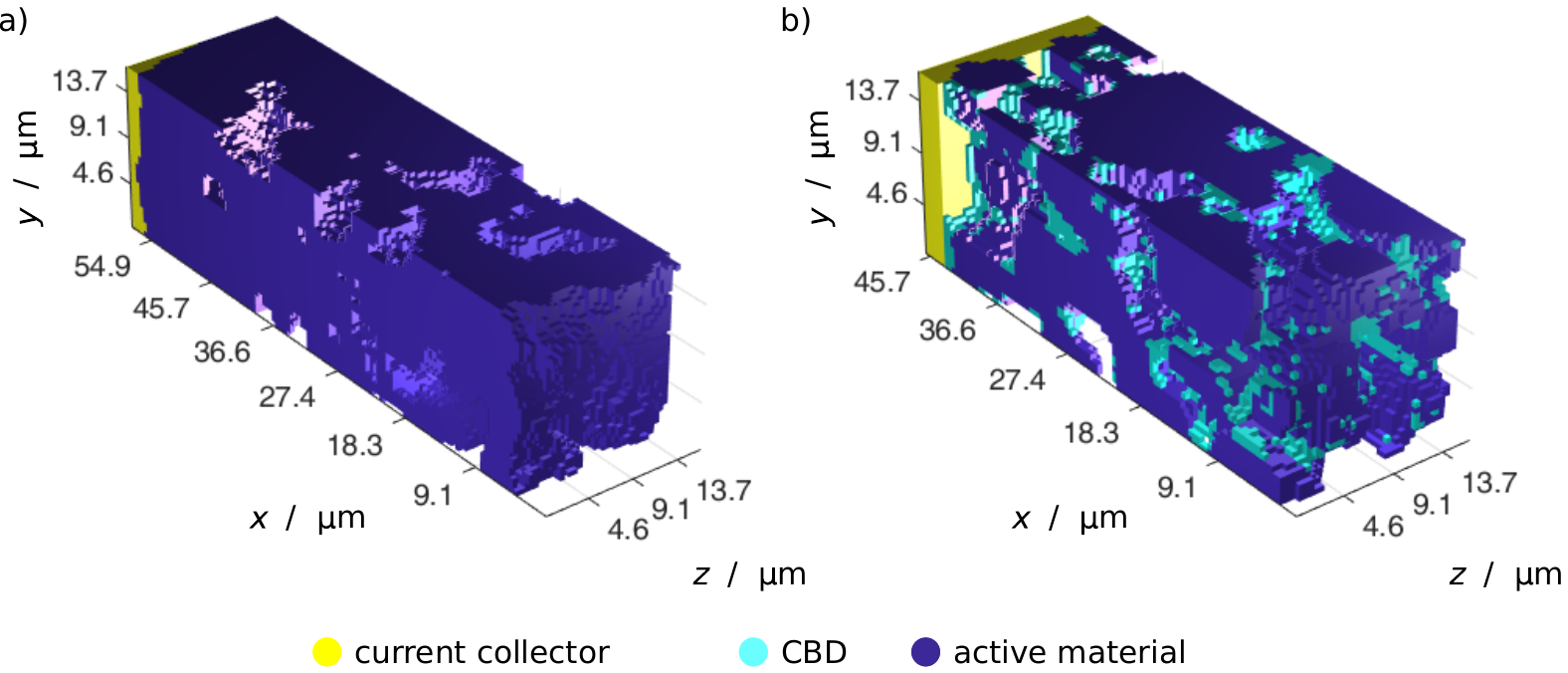}
\caption{Electrode segments of the anode a) and the cathode b) with adapted resolution, used in the BEST simulations.}
\label{pic_3d_electrodes}
\end{center}
\end{figure*}
\section{Simulations}
\label{simulations}
\subsection{Methods}
\label{Methods_Simulations}
To understand the details of the deviation in Li plating behavior, we numerically investigate the cells using the physics-based Battery and Electrochemistry Simulation Tool~(BEST)~\cite{ITWM2020}. BEST is a finite volume tool, which allows for three-dimensional, microstructure resolved simulations of electrochemical transport processes in the electrodes and the electrolyte. The tool is based on the transport theory from Latz et al.~\cite{Latz2011, Latz2015}, which describes the temporal evolution of the cell state variables~(electric and electrochemical potentials, Li$^+$ ion concentrations, ...) by calculating the current and particle fluxes between the finite volumes in a thermodynamically consistent framework~(see SI Section~S2.1~\cite{Latz2011, Latz2015, Latz2013, Doyle1993, Fuller1994}). Thus, combining BEST with highly-resolved electrode geometries~(see SI Section~S3.1~\cite{Gostick2019}) and accurate parameters of the corresponding cell materials~(see SI Section~S3.2~\cite{Nguyen2019, Lehnert2025, Valoen2005, Doyle1996, Langer2013, Kremer2020, McClelland2023, Wang2018, Sturm2019, Krause2021, Safari2011, Hein2018, Lindner2024, Landesfeind2016}) yields a detailed and precise prediction of the cell performance. Note that electrochemical characterization of carbonate-based electrolytes as employed in the LG HG2 cell is challenging and can yield inconsistent results. Interphasial effects between the bulk electrolyte and the Li metal electrodes distort the measured response in commonly used polarization experiments with symmetrical Li~|~electrolyte~|~Li cells, thereby hindering reliable evaluation of the desired parameters~\cite{Bergstrom2021, Talian2019, Lehnert2024}. To avoid these experimental limitations, we incorporate partially numerically derived parameters into our model~\cite{Lehnert2025}.\\
To capture cell degradation, we simulate the formation of the solid-electrolyte interphase~(SEI) on the active material of the anode. For this, we use an adapted version of the SEI model from von Kolzenberg et al.~\cite{Kolzenberg2020}, which has previously been described in Ref.~\cite{Bolay2022}. The derivation and the parametrization of the model can be found in SI Sections~S2.2~\cite{Kolzenberg2020, Bolay2022,Newman2019,Single2018} and~S3.3~\cite{Janakiraman2020, Single2018, Zhuang2005, Borodin2006}. Note that the SEI is described using a single parameter set for all simulations at a given temperature. This modeling approach is based on previous studies, which have shown that a unified parameter set can capture both calendar aging and cycling-induced SEI growth with good agreement to experimental data over a broad temperature range, particularly in the elevated temperature regime relevant to the preconditioning conditions considered here~\cite{Kolzenberg2022a, Philipp2026}. Based on this, we assume that the same parametrization provides a reasonable description of SEI behavior also at $-20$\,\textdegree C. \\ 
SEI growth not only contributes to cell degradation but also influences the deposition of metallic Li on the anode surface. The formation of metallic Li consumes Li$^+$ ions of the electrolyte and electrons of the anode,
\begin{equation}
\text{Li}^{+} + e^{-} \rightleftharpoons \text{Li}.
\end{equation}
The reaction is thermodynamically driven by the Li plating overpotential $\eta_\mathrm{plating}$, which measures the difference in the corresponding electrochemical potentials~\cite{Hein2016, Hein2020}. We express $\eta_\mathrm{plating}$ in Eq.~\ref{eq_overpot_plating} using the electric potential of the anode $\phi_\mathrm{s}$ and the electrochemical potential of the electrolyte $\varphi_\mathrm{e}$, both with respect to metallic Li. Since the anode surface is covered with SEI, the electrons have to traverse the SEI before participating in Li plating. This reduces their electrochemical potential by $U_\mathrm{SEI}$~(see SI Eq.~S15). The chemical potential of the electrons is incorporated in $\phi_\mathrm{s}$,
\begin{equation}
\label{eq_overpot_plating}
\eta_\mathrm{plating} = \phi_\mathrm{s} - \varphi_\mathrm{e} - U_\mathrm{SEI}.
\end{equation}
If the overpotential $\eta_\mathrm{plating}$ becomes negative, the plating condition is met and metallic Li can deposit on the anode,
\begin{equation}
\eta_\mathrm{plating} < 0.
\end{equation}
Although Li plating is not explicitly modeled in this work, we track $\eta_\mathrm{plating}$ throughout our simulations. This allows for predicting and comparing the Li plating behavior in the examined cells.\\

In analogy to the experiments, we simulate three identical cells, which follow the same protocols as described in Section~\ref{Methods_Experiments}. These protocols involve either high-temperature cycling~(cell A$_\mathrm{sim}$), high-temperature storage~(cell B$_\mathrm{sim}$), or no preconditioning~(cell C$_\mathrm{sim}$), before subsequent low-temperature cycling. The simulated cells are initialized at cell potentials of $U = 3.9\,$V~(cell A$_\mathrm{sim}$) and $U = 3.5\,$V~(cells B$_\mathrm{sim}$ and C$_\mathrm{sim}$), consistent with the experimental conditions. The influence of the initial cell potential on Li plating behavior is examined in more detail in the supplementary information~(see SI Section~S6). In the simulations, we focus on $n=2$ consecutive low-temperature cycling sets of $N=20$ cycles each, tracking the Li plating overpotential to assess the onset and relative propensity for Li plating.\\ 
All simulations employed small representative electrode segments extracted from the CT images, using adapted spatial resolutions~(see Fig.~\ref{pic_3d_electrodes} and SI Section~S3.1). Periodic boundary conditions were applied perpendicular to the through direction of the cell~($x$-axis), and all transport was modeled isothermally. The applied current density was scaled by the ratio between the current collector area in the real cells~\cite{Krause2021} and that in the simulation domain.

\subsection{Results}
\label{sim_results}
\begin{figure}[tb]
\begin{center}
\includegraphics[width=0.85\linewidth]{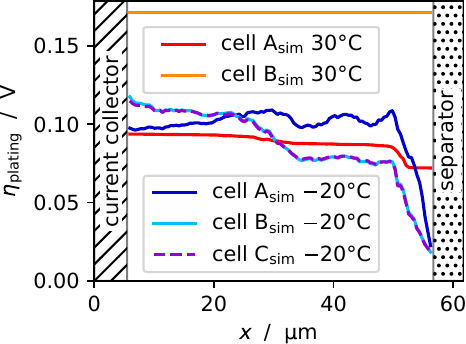}
\caption{Li plating overpotentials $\eta_\mathrm{plating}$ of cells A$_\mathrm{sim}$, B$_\mathrm{sim}$, and C$_\mathrm{sim}$ at the end of the first charge or during storage at 30\,\textdegree C and $-20$\,\textdegree C over the through direction of the anode.}
\label{eta_plating}
\end{center}
\end{figure}
The Li plating overpotential $\eta_\mathrm{plating}$ remains consistently positive during the high-temperature preconditioning and the subsequent low-temperature cycling for all three cells A$_\mathrm{sim}$, B$_\mathrm{sim}$, and C$_\mathrm{sim}$, and thus, yields no clear indications for Li plating. However, at low temperatures, $\eta_\mathrm{plating}$ reaches small values close to the separator, particularly during the first charge. This observation is consistent with the findings of Hein et al., who demonstrated through microstructure resolved cell level simulations that the Li plating reaction most likely initiates at the anode-separator-interface~\cite{Hein2016}. \\
Figure~\ref{eta_plating} shows the overpotentials averaged over the $yz$-planes, yielding a one-dimensional profile along the through direction of the cell~($x$-axis) at the end of the first charge, when $\eta_\mathrm{plating}$ reaches its minimum values. As evident in the figure, cells B$_\mathrm{sim}$ and C$_\mathrm{sim}$ exhibit very similar local Li plating overpotentials, both lower than those of cell A$_\mathrm{sim}$. Across the ten surface voxel layers of the anode's active material adjacent to the separator~($\simeq 4.57\,$\textmu m), the difference in Li plating overpotential between cells A$_\mathrm{sim}$ and B$_\mathrm{sim}$ (or C$_\mathrm{sim}$) ranges from 4\,mV to 38\,mV. While the absolute difference is moderate, using $\eta_\mathrm{plating}$ as a relative indicator still suggests that cells B$_\mathrm{sim}$ and C$_\mathrm{sim}$ are more prone to Li plating than cell A$_\mathrm{sim}$. This trend matches the experimental findings~(see Section~\ref{exp_battery_cycling}).\\ 
To identify the origin of the deviation in Li plating propensity between the cells, we analyze the effects of the different high-temperature preconditioning protocols on the time evolution of key state variables during low-temperature cycling.

\subsubsection{High-Temperature Preconditioning Process}
\begin{figure*}[t]
\begin{center}
\includegraphics[width=\linewidth]{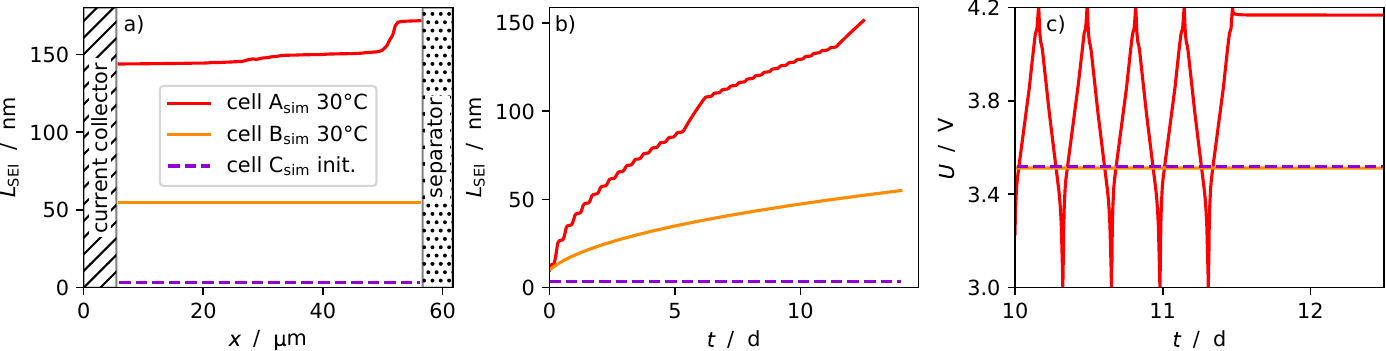}
\caption{a) SEI profile over the through direction of the anode, b) average SEI thickness~$L_\mathrm{SEI}$, and c) cell potential $U$ at the end of the high-temperature preconditioning protocols. Note that cell C$_\mathrm{sim}$ undergoes no preconditioning process, but its values are included for comparison of the cells’ initial conditions at the start of low-temperature cycling.}
\label{form_prot}
\end{center}
\end{figure*}
The cycling and storage processes at 30\,\textdegree C induce distinct differences between cells A$_\mathrm{sim}$ and B$_\mathrm{sim}$, as well as the non-preconditioned cell C$_\mathrm{sim}$ (see Fig.~\ref{form_prot}). Consequently, all three cells enter the subsequent low-temperature cycling in different initial states.\\
Firstly, the SEI grows to different thicknesses $L_\mathrm{SEI}$ over time $t$. In cell A$_\mathrm{sim}$, the SEI growth is driven by the applied overpotentials and the corresponding intercalation currents during cycling. These currents are non-uniformly distributed across the active surface and are predominantly focused on the anode regions adjacent to the separator, due to the geometrical and electrochemical properties of the anode and the electrolyte~(see SI Section~S4). Therefore, the anode of cell A$_\mathrm{sim}$ exhibits thinner SEI layers on the inner regions next to the current collector and thicker SEI layers on the outer regions next to the separator. In contrast, the calendar aging of cell B$_\mathrm{sim}$ results in a uniform distribution of the SEI across the active anode surface, and its average SEI thickness remains lower due to the low half-cell open circuit potential~(OCP) of the anode. Cell C$_\mathrm{sim}$, which undergoes no high-temperature preconditioning, is initialized with an even lower uniform SEI thickness of 10\,nm.\\ 
Secondly, the cells differ in their state-of-charge~(SOC). Cell A$_\mathrm{sim}$ is charged to the cut-off voltage of $U=4.2\,$V at the end of each preconditioning cycling set, while cell B$_\mathrm{sim}$ maintains its initial cell potential of $U=3.5\,$V during storage. Cell C$_\mathrm{sim}$, without any preconditioning, is initialized at $U=3.5\,$V~(see Sections~\ref{Methods_Experiments} and~\ref{Methods_Simulations}). This results in distinct Li$^+$ ion concentrations within the electrodes of cell A$_\mathrm{sim}$ compared to those of cells B$_\mathrm{sim}$ and C$_\mathrm{sim}$ at the beginning of the low-temperature cycling. Note that the elevated SOC of cell A$_\mathrm{sim}$ during the multi-hour rest periods between and after the preconditioning cycling sets leads to pronounced calendar aging, which explains the irregular SEI~growth observed in Figure~\ref{form_prot}b. 

\subsubsection{Low-Temperature Cycling}
After completion of the respective preconditioning process, all three cells A$_\mathrm{sim}$, B$_\mathrm{sim}$, and C$_\mathrm{sim}$ are cooled and numerically cycled at $-20$\,\textdegree C. As mentioned above, the cells exhibit low but deviating Li plating overpotentials at the anode regions adjacent to the separator at this temperature, particularly at the end of the first charge~(see Fig.~\ref{eta_plating}). To understand the origins of the observed phenomena, we average key state variables across the ten surface voxel layers of the anode's active material~($\simeq 4.57\,$\textmu m) located next to the separator and analyze their time evolution.\\

\begin{figure*}[t!]
\begin{center}
\includegraphics[width=\linewidth]{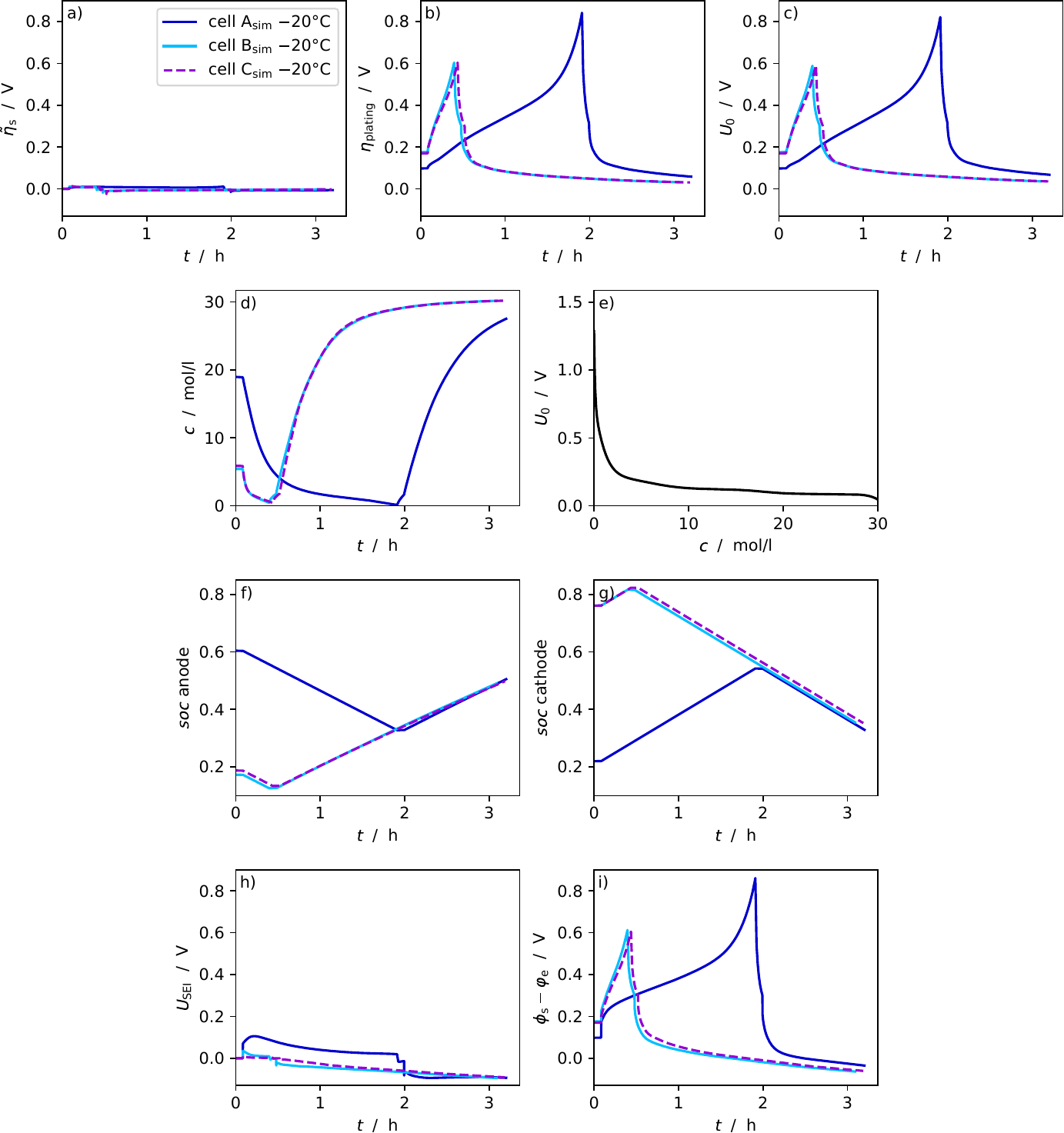}
\caption{Various cell state variables during the first low-temperature discharge and charge process: a) Butler-Volmer overpotential $\tilde{\eta}_\mathrm{s}$, b) Li plating overpotential $\eta_\mathrm{plating}$, c) OCP $U_0$ of the anode, d) Li$^+$ ion concentration $c$ of the anode, e) OCP $U_0$ of the anode as a function of~$c$~\cite{Hein2018}, overall SOC $\left(\mathit{soc}=c/c_\mathrm{max}\right)$ of the anode f) and cathode g), h) potential drop due to the SEI $U_\mathrm{SEI}$, and i) potential difference between the electrical potential of the anode $\phi_\mathrm{s}$ and the electrochemical potential of the electrolyte $\varphi_\mathrm{e}$. Quantities shown in a)--d) and h)--i) are averaged across the ten anode voxel layers adjacent to the separator. The SOC of each electrode is averaged over its entire volume.}
\label{c_-20C}
\end{center}
\end{figure*}
\textit{First Cycle}\\
Figure~\ref{c_-20C}a--c show the local average of the Butler-Volmer overpotential $\tilde{\eta}_\mathrm{s}$, the Li plating overpotential $\eta_\mathrm{plating}$ and the OCP $U_0$ of the anode during the first discharge and charge. These potentials are linked via the Butler-Volmer equation describing the intercalation currents on the active surface material~(see SI Eqs.~S5--S9).
\begin{equation}
\tilde{\eta}_\mathrm{s} = \eta_\mathrm{plating} - U_0
\end{equation}
$\tilde{\eta}_\mathrm{s}$ shows similarly positive and negative values during discharge and charge in all three cells~(see Fig.~\ref{c_-20C}a). However, it remains relatively small and varies only slightly over time, particularly during the individual charge and discharge processes. In contrast, the Li plating overpotential $\eta_\mathrm{plating}$ and the OCP $U_0$ change monotonically, spanning significantly larger ranges~(see Fig.~\ref{c_-20C}b,c). This indicates a strong correlation between the Li plating overpotential and the OCP.\\
The observed correlation implies that differences in Li plating propensity between the cells can be attributed to deviations in their OCPs, which depend solely on the Li$^+$ ion concentration of the anode. Therefore, to understand the Li plating propensity in the cells, we have to examine the evolution of their Li$^+$ ion concentration in the anode regions adjacent to the separator.\\

The low-temperature cycling begins with a discharge followed by a charge, during which the local Li$^+$ ion concentration in the anode decreases and subsequently increases in all three cells~(see Fig.~\ref{c_-20C}d). However, the time evolution of the concentrations differs.\\
Cell A$_\mathrm{sim}$ starts cycling with a higher cell SOC induced by the high-temperature preconditioning protocol. Therefore, it discharges for longer times than cells B$_\mathrm{sim}$ and C$_\mathrm{sim}$, transferring overall more Li$^+$ ions from the anode into the cathode. Due to the low-temperature kinetics, the ions are primarily deintercalated from the anode regions adjacent to the separator~(see SI Section~S4). This results in a strong concentration gradient in through direction of the anode, with elevated concentrations near the current collector and depleted concentrations near the separator. As the discharge progresses, the concentrations near the separator become so low that the corresponding OCP increases non-linearly~\cite{Hein2018}~(see Fig.~\ref{c_-20C}e), hindering further deintercalation. Therefore, the discharge cut-off voltage is reached prematurely, where the anode and cathode still exhibit relatively high and low overall concentrations, respectively~(see Fig.~\ref{c_-20C}f,g). This shortens the duration of the subsequent charge, reducing the amount of charge transferred from the cathode back into the anode. Although the local Li$^+$ ion concentration in the anode regions adjacent to the separator still increases during charge, it remains relatively low at the end of the process. This leads to elevated OCPs and, due to the correlation mentioned above, to relatively high Li plating overpotentials.\\
Cells B$_\mathrm{sim}$ and C$_\mathrm{sim}$ show almost identical time evolutions of the anode Li$^+$ ion concentration, despite their initial difference in uniform SEI thickness. In contrast to cell A$_\mathrm{sim}$, both cells start cycling at a lower cell SOC and thus, discharge for a shorter duration. This generates a less pronounced concentration gradient within the anode and consequently, results in a reduced influence of the non-linear OCP increase on the cell potential. Therefore, both cells B$_\mathrm{sim}$ and C$_\mathrm{sim}$ reach an overall lower SOC at the end of their discharge. As a result, the subsequent charge duration is prolonged, inducing elevated local Li$^+$ ion concentrations in the anode regions adjacent to the separator. This results in lower OCPs and hence, reduced overpotentials for Li plating at the end of charge.\\ 

These observations suggest that the primary difference between the cells, induced by the high-temperature preconditioning process, lies in the deviations of their initial SOCs at the beginning of the low-temperature cycling. The variations in SEI profile and thickness have only negligible effects during the first discharge and charge process. Although the SEI theoretically influences the intercalation currents and the time evolution of the concentrations, our simulations exhibit relatively small potential drops $U_\mathrm{SEI}$ during this period, especially in comparison to the dominant contribution of the Li plating overpotential $\phi_\mathrm{s} - \varphi_\mathrm{e}$~(see Fig.~\ref{c_-20C}h,i). Additionally, $U_\mathrm{SEI}$ is nearly identical for all three cells. Therefore, the observed difference in Li plating propensity can be primarily attributed to the initial SOC deviation rather than to differences in the SEI.\\
We verify this conclusion by conducting two additional low-temperature cycling simulations, in which cells A$_\mathrm{sim}$ and B$_\mathrm{sim}$ are resimulated using swapped SEIs -- that is, cell A$_\mathrm{sim}$ adopts the SEI of cell B$_\mathrm{sim}$ formed at the end of its high-temperature preconditioning process and vice versa~(see SI Section~S5). These modified simulations allow for a SOC-matched comparison to the original simulations, confirming that the differences in Li plating overpotentials at the end of the first charge are dominated by the induced SOC rather than the SEI. \\ 
Additional analysis on the influence of varying uniform SEI thicknesses and the overall cell SOCs on the Li plating overpotential during the first discharge and charge process is provided in the supplementary information~(see SI Section~S6).\\ 

\begin{figure*}[bt]
\begin{center}
\includegraphics[width=\linewidth]{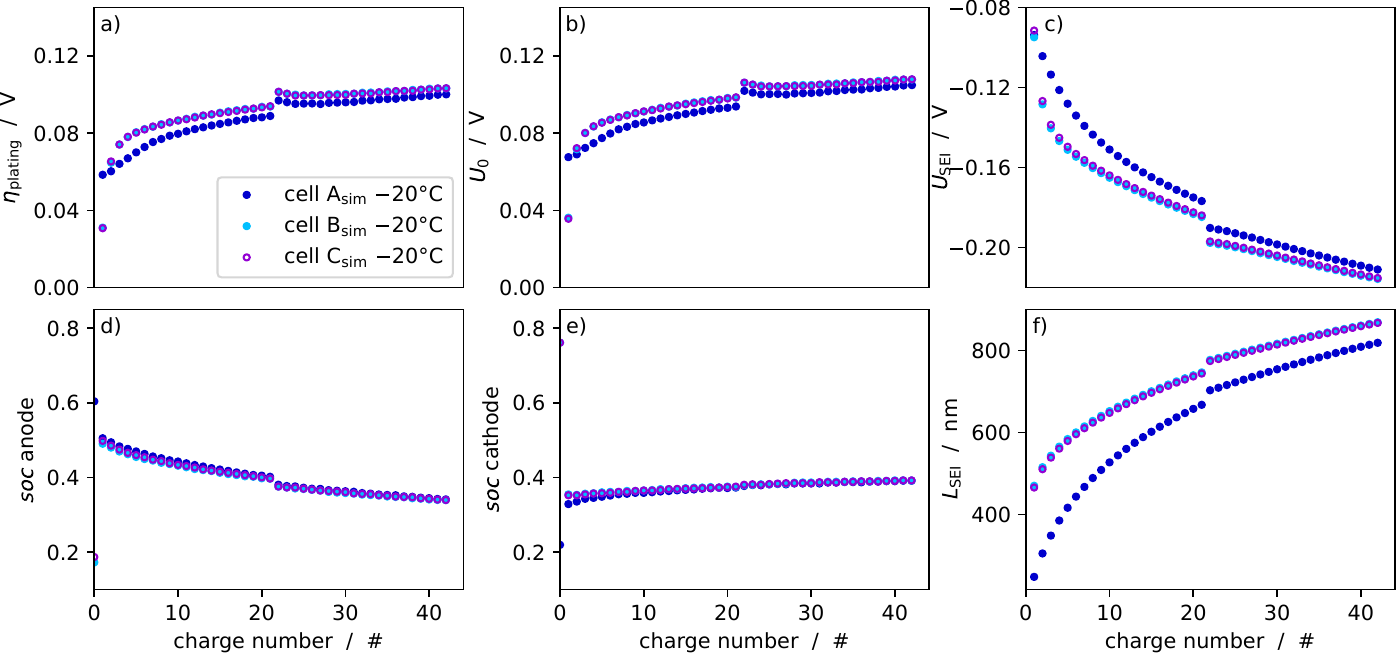}
\caption{a) Li plating overpotential $\eta_\mathrm{plating}$, b) OCP $U_0$ of the anode, c) potential drop due to the SEI $U_\mathrm{SEI}$, overall SOCs of anode d) and cathode e), and, f) SEI thickness $L_\mathrm{SEI}$ at the end of each charge. The quantities shown in a)--c) and f) are averaged across the ten anode voxel layers adjacent to the separator. The data points in d) and e) at the 0th charge number correspond to the SOCs at the beginning of the low-temperature cycling.}
\label{time_evolution_-20C}
\end{center}
\end{figure*}
\textit{Subsequent Cycles}\\
During the subsequent cycles, the local Li plating overpotentials of the cells A$_\mathrm{sim}$, B$_\mathrm{sim}$, and C$_\mathrm{sim}$ increase slightly and their deviation diminishes with increasing cycle number~(see Fig.~\ref{time_evolution_-20C}). The origin of this phenomenon lies in increased internal resistances.\\
At the end of the first charge, cell A$_\mathrm{sim}$ reaches a reduced cell SOC compared to its initial state at the beginning of the low-temperature cycling, due to the impeded low-temperature transport. Therefore, it starts its second cycle with electrode SOCs similar to those of cells B$_\mathrm{sim}$ and C$_\mathrm{sim}$, effectively eliminating significant deviations between the OCPs of the cells and thus, their Li plating overpotentials at the end of each charge process. Consequently, increased initial cell SOCs suppress the Li plating propensity only during the first charge in our simulations.\\
Additionally, the SEI grows over the cycles, further increasing the amplitude of the potential drop $U_\mathrm{SEI}$. This elevates the internal resistance of the cells and shortens the duration of subsequent charge and discharge processes, as the cut-off voltages are reached earlier. Therefore, less charge is transferred between the electrodes, leading to less extreme concentrations and OCPs in the anode regions adjacent to the separator. Consequently, the Li plating overpotentials increase with increasing cycle number. This effect is slightly more pronounced in cells B$_\mathrm{sim}$ and C$_\mathrm{sim}$, which develop locally thicker SEI compared to cell A$_\mathrm{sim}$.\\
Although this trend would suggest that the cells should become less prone to Li plating at higher cycle numbers, experimental observations have shown the opposite: cells B$_\mathrm{exp}$ and C$_\mathrm{exp}$ displayed increasing amounts of stripped Li $Q_\mathrm{stripped}$ with cycling, starting between the second and seventh discharge, depending on the cell~(see Fig.~\ref{pic_CE}). While the simulations can explain the reduction in $Q_\mathrm{stripped}$ during the initial cycles, they cannot account for the increase observed in subsequent cycles. This discrepancy indicates that Li plating during the initial cycles modifies the cells in a way that promotes further plating -- an effect not captured by simulations that do not model plating and its consequences.

\section{Discussion}
\label{discussion}
In the previous Sections, we experimentally and numerically investigated the influence of different high-temperature protocols on Li plating during subsequent low-temperature cycling in the LG HG2 cell. Both the experiments and the simulations show the same trends. The first protocol -- cycling at 30\,\textdegree C -- makes the cell more resistant against Li plating during cycling at $-20$\,\textdegree C than the second or third protocol, which involve storage at 30\,\textdegree C and direct cycling at $-20$\,\textdegree C without prior preconditioning, respectively. The simulations suggest that the difference in Li plating behavior originates from differences in the respective cell SOCs induced by the different preconditioning protocols. Thereby, lower initial cell SOCs result in locally elevated Li$^+$ ion concentrations in the anode at the end of the first charge, which promotes Li plating. This aligns with literature findings, indicating that metallic Li deposition is more likely to occur with increasing SOC~\cite{Janakiraman2020, Waldmann2018}.\\ 
However, in contrast to the experiments, the simulations exhibit throughout positive Li plating overpotentials, implying that Li plating is thermodynamically unfeasible. Accordingly, the simulations capture only relative differences in Li plating propensity rather than predicting absolute Li plating occurrence. This discrepancy with the experimental findings could be due to effects, which are not yet captured by our cell model or its parametrization.\\
Besides possible inaccuracies in the parametrization of the electrolyte, the separator and the SEI~(see SI Sections~S3.2 and~S3.3), our SEI model simulates the SEI thickness virtually, i.e.\ the SEI does not occupy any volume in the anode geometry. Therefore, while the SEI influences the occurring potentials in the cell, it does not modify the porosity or tortuosity through pore clogging.\\
Additionally, our SEI model does not distinguish between different SEI components. The literature often describes the SEI as a mixture of organic and inorganic components, which form an inner compact layer and a porous outer layer, or even more complex structures~\cite{Lu2014, Kolzenberg2022, Peled2017, Huang2019, An2016}. While the SEI formation mechanisms are still debated, it is generally observed that different components form at different anode potentials~vs.~Li/Li$^+$~\cite{Lu2014, An2016}. Therefore, the cell undergoing high-temperature cycling could form a differently composed SEI, which is more stable against and already adapted to volumetric changes of the anode during the low-temperature cycling, compared to the other cells. In contrast, the SEI of both the stored cell and the cell directly cycled at $-20$\,\textdegree C may be more prone to cracking, particularly during the first cycle. Such cracking locally removes any passivating SEI layer from the anode surface, placing the active materials in direct contact with the electrolyte. As a result, the potential drop $U_\mathrm{SEI}$ vanishes and the Li plating overpotential is solely determined by the potential difference $\phi_\mathrm{s} - \varphi_\mathrm{e}$~(see Eq.~\ref{eq_overpot_plating}). Since $\phi_\mathrm{s} - \varphi_\mathrm{e}$ reaches negative values during charge in the anode regions adjacent to the separator~(see Fig.~\ref{c_-20C}i), any crack occurring within this region would therefore result in negative Li plating overpotentials, making the deposition of metallic Li thermodynamically possible.\\ 
Finally, our simulations represent an idealized scenario and therefore do not account for potential inhomogeneous coverage of the anode by the CBD or incomplete particle contacting of the active particles. Both effects could further lower the overpotential for Li plating and thus promote the onset of metallic Li deposition~\cite{Traskunov2021, Traskunov2021a, Traskunov2022}.

\section{Conclusion}
In this work, we experimentally and numerically investigated the effects of different high-temperature preconditioning protocols on Li plating behavior during subsequent low-temperature cycling in the COTS LG HG2 cell. The first preconditioning protocol involved two-week cycling at 30\,\textdegree C, the second protocol consisted of cell storage under identical conditions, while the third protocol involved no preconditioning and direct cycling at low temperature. All cycles were performed at a constant current corresponding to C/5.\\
Experimental cycling tests revealed that the different protocols result in distinct low-temperature Li plating behavior. Cells stored at high temperature or directly cycled at low temperature showed evidence of Li plating during operation at $-20\,$\textdegree C, whereas the cells subjected to high-temperature cycling showed no such indications.\\
These experimental findings are consistent with the trends observed in our three-dimensional, microstructure resolved simulations, which replicated the experiments. The simulations revealed lower Li plating overpotentials -- and thus a higher plating tendency -- in the stored cell and the non-preconditioned cell compared to the cycled cell. This deviation primarily originates from differences in the cell state of charge induced by the preconditioning protocols, rather than from induced variations in the SEI.\\
Our work demonstrates that combining experimental testing with detailed cell level simulations provides deep insights into the underlying processes and degradation mechanisms by accessing key state variables that would otherwise remain hidden. Such insights are particularly valuable for predicting cell performance in space applications, where batteries are exposed to extreme conditions and any failure can lead to severe risks and costs. Therefore, continued advancement of experimental methods and numerical models is essential to enable the design of reliable batteries for future space missions with rigorous mission requirements.

\appendix
\section*{Acknowledgements}
We thank the project DeepBat, a collaboration between the National Aeronautics and Space Administration -- Jet Propulsion Laboratory~(NASA -- JPL) and the German Aerospace Center~(DLR), for their support, and gratefully acknowledge the support by the state of Baden-Württemberg through bwHPC and the German Research Foundation~(DFG) through grant no INST 40/575-1 FUGG~(JUSTUS 2 cluster). Further appreciation is extended to the Center for Electrochemical Energy Storage Ulm \& Karlsruhe~(CELEST).

\section*{Author Contributions}
Lukas~Lehnert contributed to the conceptualization of the paper, drafted the manuscript and prepared the figures for both the manuscript and the supplementary information. He parametrized and conducted the simulations, and analyzed the results. Dr.~Keith~B.~Chin, Dr.~Frederick~C.~Krause and John~P.~Ruiz performed the high-temperature preconditioning and low-temperature cycling of the LG HG2 cell. Dr.~Simon~Hein binarized the CT data and supervised the simulations. John~Bescup and Gil~Garteiz imaged the electrode microstructures using CT and contributed to the binarization of the obtained data. Prof.~Dr.~Arnulf Latz contributed to the conceptualization of the paper and to the acquisition of funding. Dr.~Erik~J.~Brandon and Prof.~Dr.~Birger~Horstmann supervised the experimental and numerical work, respectively, contributed to the conceptualization of the paper, reviewed the article, and acquired funding.

\section*{Conflict of Interest}
The authors declare no conflict of interest.

\bibliographystyle{elsarticle-num} 
\bibliography{Paper3.bib}

@Article{Latz2011,
  author    = {Latz, A. and Zausch, J.},
  journal   = {Journal of Power Sources},
  title     = {Thermodynamic consistent transport theory of Li-ion batteries},
  year      = {2011},
  issn      = {0378-7753},
  month     = mar,
  number    = {6},
  pages     = {3296--3302},
  volume    = {196},
  doi       = {10.1016/j.jpowsour.2010.11.088},
  publisher = {Elsevier BV},
}

@Article{Latz2015,
  author    = {Latz, Arnulf and Zausch, Jochen},
  journal   = {Beilstein Journal of Nanotechnology},
  title     = {Multiscale modeling of lithium ion batteries: thermal aspects},
  year      = {2015},
  issn      = {2190-4286},
  month     = apr,
  pages     = {987--1007},
  volume    = {6},
  doi       = {10.3762/bjnano.6.102},
  publisher = {Beilstein Institut},
}

@Article{Kolzenberg2020,
  author    = {von Kolzenberg, Lars and Latz, Arnulf and Horstmann, Birger},
  journal   = {ChemSusChem},
  title     = {Solid–Electrolyte Interphase During Battery Cycling: Theory of Growth Regimes},
  year      = {2020},
  issn      = {1864-564X},
  month     = jun,
  number    = {15},
  pages     = {3901--3910},
  volume    = {13},
  doi       = {10.1002/cssc.202000867},
  publisher = {Wiley},
}

@Article{Bolay2022,
  author    = {Bolay, Linda J. and Schmitt, Tobias and Hein, Simon and Mendoza-Hernandez, Omar S. and Hosono, Eiji and Asakura, Daisuke and Kinoshita, Koichi and Matsuda, Hirofumi and Umeda, Minoru and Sone, Yoshitsugu and Latz, Arnulf and Horstmann, Birger},
  journal   = {Journal of Power Sources Advances},
  title     = {Microstructure-resolved degradation simulation of lithium-ion batteries in space applications},
  year      = {2022},
  issn      = {2666-2485},
  month     = mar,
  pages     = {100083},
  volume    = {14},
  doi       = {10.1016/j.powera.2022.100083},
  publisher = {Elsevier BV},
}

@Article{Single2018,
  author    = {Single, Fabian and Latz, Arnulf and Horstmann, Birger},
  journal   = {ChemSusChem},
  title     = {Identifying the Mechanism of Continued Growth of the Solid–Electrolyte Interphase},
  year      = {2018},
  issn      = {1864-564X},
  month     = apr,
  number    = {12},
  pages     = {1950--1955},
  volume    = {11},
  doi       = {10.1002/cssc.201800077},
  publisher = {Wiley},
}

@Article{ITWM2020,
  author  = {ITWM},
  journal = {http://itwm.fraunhofer.de/best},
  title   = {Best - battery and electrochemistry simulation tool},
  year    = {2020},
}

@Book{Newman2019,
  author    = {Newman, John and Balsara, Nitash P.},
  publisher = {Wiley \& Sons, Incorporated, John},
  title     = {Electrochemical Systems},
  year      = {2019},
  isbn      = {9781119514602},
  pages     = {704},
}

@Article{Krause2021,
  author    = {Krause, F. C. and Ruiz, J. P. and Jones, S. C. and Brandon, E. J. and Darcy, E. C. and Iannello, C. J. and Bugga, R. V.},
  journal   = {Journal of The Electrochemical Society},
  title     = {Performance of Commercial Li-Ion Cells for Future NASA Missions and Aerospace Applications},
  year      = {2021},
  issn      = {1945-7111},
  month     = apr,
  number    = {4},
  pages     = {040504},
  volume    = {168},
  doi       = {10.1149/1945-7111/abf05f},
  publisher = {The Electrochemical Society},
}

@Article{Nguyen2019,
  author    = {Nguyen, Thi Thu Dieu and Abada, Sara and Lecocq, Amandine and Bernard, Julien and Petit, Martin and Marlair, Guy and Grugeon, Sylvie and Laruelle, Stéphane},
  journal   = {World Electric Vehicle Journal},
  title     = {Understanding the Thermal Runaway of Ni-Rich Lithium-Ion Batteries},
  year      = {2019},
  issn      = {2032-6653},
  month     = nov,
  number    = {4},
  pages     = {79},
  volume    = {10},
  doi       = {10.3390/wevj10040079},
  publisher = {MDPI AG},
}

@Article{Valoen2005,
  author    = {Val{\o}en, Lars Ole and Reimers, Jan N.},
  journal   = {Journal of The Electrochemical Society},
  title     = {Transport Properties of LiPF$_6$-Based Li-Ion Battery Electrolytes},
  year      = {2005},
  issn      = {0013-4651},
  number    = {5},
  pages     = {A882},
  volume    = {152},
  doi       = {10.1149/1.1872737},
  publisher = {The Electrochemical Society},
}

@Article{Safari2011,
  author    = {Safari, M. and Delacourt, C.},
  journal   = {Journal of The Electrochemical Society},
  title     = {Modeling of a Commercial Graphite/LiFePO4 Cell},
  year      = {2011},
  issn      = {0013-4651},
  number    = {5},
  pages     = {A562},
  volume    = {158},
  doi       = {10.1149/1.3567007},
  publisher = {The Electrochemical Society},
}

@Article{Langer2013,
  author    = {Langer, J. and Epp, V. and Heitjans, P. and Mautner, F. A. and Wilkening, M.},
  journal   = {Physical Review B},
  title     = {Lithium motion in the anode material LiC$_6$ as seen via time-domain $\prescript{7}{}{\mathrm{Li}}$ NMR},
  year      = {2013},
  issn      = {1550-235X},
  month     = sep,
  number    = {9},
  pages     = {094304},
  volume    = {88},
  doi       = {10.1103/physrevb.88.094304},
  publisher = {American Physical Society (APS)},
}

@PhdThesis{Hein2018,
  author    = {Hein, Simon},
  title     = {Modeling of lithium plating in lithium-ion-batteries},
  year      = {2018},
  copyright = {Standard},
  doi       = {10.18725/OPARU-7939},
  language  = {en},
  publisher = {Universität Ulm},
}

@Article{Sturm2019,
  author    = {Sturm, J. and Rheinfeld, A. and Zilberman, I. and Spingler, F.B. and Kosch, S. and Frie, F. and Jossen, A.},
  journal   = {Journal of Power Sources},
  title     = {Modeling and simulation of inhomogeneities in a 18650 nickel-rich, silicon-graphite lithium-ion cell during fast charging},
  year      = {2019},
  issn      = {0378-7753},
  month     = feb,
  pages     = {204--223},
  volume    = {412},
  doi       = {10.1016/j.jpowsour.2018.11.043},
  publisher = {Elsevier BV},
}

@Article{McClelland2023,
  author    = {McClelland, Innes and Booth, Samuel G. and Anthonisamy, Nirmalesh N. and Middlemiss, Laurence A. and Pérez, Gabriel E. and Cussen, Edmund J. and Baker, Peter J. and Cussen, Serena A.},
  journal   = {Chemistry of Materials},
  title     = {Direct Observation of Dynamic Lithium Diffusion Behavior in Nickel-Rich, LiNi$_{0.8}$Mn$_{0.1}$Co$_{0.1}$O$_2$ (NMC811) Cathodes Using Operando Muon Spectroscopy},
  year      = {2023},
  issn      = {1520-5002},
  month     = may,
  number    = {11},
  pages     = {4149--4158},
  volume    = {35},
  doi       = {10.1021/acs.chemmater.2c03834},
  publisher = {American Chemical Society (ACS)},
}

@Article{Wang2018,
  author    = {Wang, Shanyu and Yan, Mengyu and Li, Yun and Vinado, Carolina and Yang, Jihui},
  journal   = {Journal of Power Sources},
  title     = {Separating electronic and ionic conductivity in mix-conducting layered lithium transition-metal oxides},
  year      = {2018},
  issn      = {0378-7753},
  month     = jul,
  pages     = {75--82},
  volume    = {393},
  doi       = {10.1016/j.jpowsour.2018.05.005},
  publisher = {Elsevier BV},
}

@Article{Kremer2020,
  author    = {Kremer, Lea Sophie and Danner, Timo and Hein, Simon and Hoffmann, Alice and Prifling, Benedikt and Schmidt, Volker and Latz, Arnulf and Wohlfahrt‐Mehrens, Margret},
  journal   = {Batteries \& Supercaps},
  title     = {Influence of the Electrolyte Salt Concentration on the Rate Capability of Ultra‐Thick NCM 622 Electrodes},
  year      = {2020},
  issn      = {2566-6223},
  month     = aug,
  number    = {11},
  pages     = {1172--1182},
  volume    = {3},
  doi       = {10.1002/batt.202000098},
  publisher = {Wiley},
}

@Article{Landesfeind2016,
  author    = {Landesfeind, Johannes and Hattendorff, Johannes and Ehrl, Andreas and Wall, Wolfgang A. and Gasteiger, Hubert A.},
  journal   = {Journal of The Electrochemical Society},
  title     = {Tortuosity Determination of Battery Electrodes and Separators by Impedance Spectroscopy},
  year      = {2016},
  issn      = {1945-7111},
  number    = {7},
  pages     = {A1373--A1387},
  volume    = {163},
  doi       = {10.1149/2.1141607jes},
  publisher = {The Electrochemical Society},
}

@Article{Doyle1996,
  author    = {Doyle, Marc and Newman, John and Gozdz, Antoni S. and Schmutz, Caroline N. and Tarascon, Jean‐Marie},
  journal   = {Journal of The Electrochemical Society},
  title     = {Comparison of Modeling Predictions with Experimental Data from Plastic Lithium Ion Cells},
  year      = {1996},
  issn      = {1945-7111},
  month     = jun,
  number    = {6},
  pages     = {1890--1903},
  volume    = {143},
  doi       = {10.1149/1.1836921},
  publisher = {The Electrochemical Society},
}

@Article{Zhuang2005,
  author    = {Zhuang, Guorong V. and Xu, Kang and Yang, Hui and Jow, T. Richard and Ross, Philip N.},
  journal   = {The Journal of Physical Chemistry B},
  title     = {Lithium Ethylene Dicarbonate Identified as the Primary Product of Chemical and Electrochemical Reduction of EC in 1.2 M LiPF$_6$/EC:EMC Electrolyte},
  year      = {2005},
  issn      = {1520-5207},
  month     = aug,
  number    = {37},
  pages     = {17567--17573},
  volume    = {109},
  doi       = {10.1021/jp052474w},
  publisher = {American Chemical Society (ACS)},
}

@Article{Borodin2006,
  author    = {Borodin, Oleg and Smith, Grant D. and Fan, Peng},
  journal   = {The Journal of Physical Chemistry B},
  title     = {Molecular Dynamics Simulations of Lithium Alkyl Carbonates},
  year      = {2006},
  issn      = {1520-5207},
  month     = oct,
  number    = {45},
  pages     = {22773--22779},
  volume    = {110},
  doi       = {10.1021/jp0639142},
  publisher = {American Chemical Society (ACS)},
}

@Article{Hein2016,
  author    = {Hein, Simon and Latz, Arnulf},
  journal   = {Electrochimica Acta},
  title     = {Influence of local lithium metal deposition in 3D microstructures on local and global behavior of Lithium-ion batteries},
  year      = {2016},
  issn      = {0013-4686},
  month     = may,
  pages     = {354--365},
  volume    = {201},
  doi       = {10.1016/j.electacta.2016.01.220},
  publisher = {Elsevier BV},
}

@Article{Hein2020,
  author    = {Hein, Simon and Danner, Timo and Latz, Arnulf},
  journal   = {ACS Applied Energy Materials},
  title     = {An Electrochemical Model of Lithium Plating and Stripping in Lithium Ion Batteries},
  year      = {2020},
  issn      = {2574-0962},
  month     = jul,
  number    = {9},
  pages     = {8519--8531},
  volume    = {3},
  doi       = {10.1021/acsaem.0c01155},
  publisher = {American Chemical Society (ACS)},
}

@Article{Krause2020,
  author    = {Krause, Frederick C. and Loveland, Jessica A. and Smart, Marshall C. and Brandon, Erik J. and Bugga, Ratnakumar V.},
  journal   = {Journal of Power Sources},
  title     = {Implementation of commercial Li-ion cells on the MarCO deep space CubeSats},
  year      = {2020},
  issn      = {0378-7753},
  month     = feb,
  pages     = {227544},
  volume    = {449},
  doi       = {10.1016/j.jpowsour.2019.227544},
  publisher = {Elsevier BV},
}

@InProceedings{pearson2005small,
  author       = {Pearson, Chris and Thwaite, Carl and Russel, Nick},
  booktitle    = {Paper AIAA RS3-2005-5003 presented at the 3rd Responsive Space Conference. Los Angeles, CA, USA},
  title        = {Small cell lithium-ion batteries: the responsive solution for space energy storage},
  year         = {2005},
  organization = {Citeseer},
  pages        = {25--28},
}

@Article{Jones2022,
  author    = {Jones, John-Paul and Smart, Marshall C. and Krause, Frederick C. and West, William C. and Brandon, Erik J.},
  journal   = {Joule},
  title     = {Batteries for robotic spacecraft},
  year      = {2022},
  issn      = {2542-4351},
  month     = may,
  number    = {5},
  pages     = {923--928},
  volume    = {6},
  doi       = {10.1016/j.joule.2022.04.004},
  publisher = {Elsevier BV},
}

@Article{Chin2018,
  author    = {Chin, Keith B. and Brandon, Erik J. and Bugga, Ratnakumar V. and Smart, Marshall C. and Jones, Simon C. and Krause, Frederick C. and West, William C. and Bolotin, Gary G.},
  journal   = {Proceedings of the IEEE},
  title     = {Energy Storage Technologies for Small Satellite Applications},
  year      = {2018},
  issn      = {1558-2256},
  month     = mar,
  number    = {3},
  pages     = {419--428},
  volume    = {106},
  doi       = {10.1109/jproc.2018.2793158},
  publisher = {Institute of Electrical and Electronics Engineers (IEEE)},
}

@Article{Gupta2020,
  author    = {Gupta, Abhay and Manthiram, Arumugam},
  journal   = {Advanced Energy Materials},
  title     = {Designing Advanced Lithium‐Based Batteries for Low‐Temperature Conditions},
  year      = {2020},
  issn      = {1614-6840},
  month     = aug,
  number    = {38},
  volume    = {10},
  doi       = {10.1002/aenm.202001972},
  publisher = {Wiley},
}

@Article{Waldmann2018,
  author    = {Waldmann, Thomas and Hogg, Björn-Ingo and Wohlfahrt-Mehrens, Margret},
  journal   = {Journal of Power Sources},
  title     = {Li plating as unwanted side reaction in commercial Li-ion cells – A review},
  year      = {2018},
  issn      = {0378-7753},
  month     = apr,
  pages     = {107--124},
  volume    = {384},
  doi       = {10.1016/j.jpowsour.2018.02.063},
  publisher = {Elsevier BV},
}

@Article{Smart2011,
  author    = {Smart, M. C. and Ratnakumar, B. V.},
  journal   = {Journal of The Electrochemical Society},
  title     = {Effects of Electrolyte Composition on Lithium Plating in Lithium-Ion Cells},
  year      = {2011},
  issn      = {1945-7111},
  month     = feb,
  number    = {4},
  pages     = {A379--A389},
  volume    = {158},
  doi       = {10.1149/1.3544439},
  publisher = {The Electrochemical Society},
}

@Article{Janakiraman2020,
  author    = {Janakiraman, Umamaheswari and Garrick, Taylor R. and Fortier, Mary E.},
  journal   = {Journal of The Electrochemical Society},
  title     = {Review—Lithium Plating Detection Methods in Li-Ion Batteries},
  year      = {2020},
  issn      = {1945-7111},
  month     = dec,
  number    = {16},
  pages     = {160552},
  volume    = {167},
  doi       = {10.1149/1945-7111/abd3b8},
  publisher = {The Electrochemical Society},
}

@Article{Lindner2024,
  author    = {Lindner, Adrian and Both, Svenja and Menesklou, Wolfgang and Hein, Simon and Danner, Timo and Latz, Arnulf and Krewer, Ulrike},
  journal   = {Batteries \& Supercaps},
  title     = {Analyzing and Improving Conductive Networks in Commercial High‐Energy Ni‐rich Cathodes},
  year      = {2024},
  issn      = {2566-6223},
  month     = sep,
  doi       = {10.1002/batt.202400503},
  publisher = {Wiley},
}

@Article{Bugga2010,
  author    = {Bugga, Ratnakumar V. and Smart, Marshall C.},
  journal   = {ECS Transactions},
  title     = {Lithium Plating Behavior in Lithium-Ion Cells},
  year      = {2010},
  issn      = {1938-6737},
  month     = apr,
  number    = {36},
  pages     = {241--252},
  volume    = {25},
  doi       = {10.1149/1.3393860},
  publisher = {The Electrochemical Society},
}

@Article{Lu2014,
  author    = {Lu, Peng and Li, Chen and Schneider, Eric W. and Harris, Stephen J.},
  journal   = {The Journal of Physical Chemistry C},
  title     = {Chemistry, Impedance, and Morphology Evolution in Solid Electrolyte Interphase Films during Formation in Lithium Ion Batteries},
  year      = {2014},
  issn      = {1932-7455},
  month     = jan,
  number    = {2},
  pages     = {896--903},
  volume    = {118},
  doi       = {10.1021/jp4111019},
  publisher = {American Chemical Society (ACS)},
}

@Article{Kolzenberg2022,
  author    = {von Kolzenberg, Lars and Werres, Martin and Tetzloff, Jonas and Horstmann, Birger},
  journal   = {Physical Chemistry Chemical Physics},
  title     = {Transition between growth of dense and porous films: theory of dual-layer SEI},
  year      = {2022},
  issn      = {1463-9084},
  number    = {31},
  pages     = {18469--18476},
  volume    = {24},
  doi       = {10.1039/d2cp00188h},
  publisher = {Royal Society of Chemistry (RSC)},
}

@Article{An2016,
  author    = {An, Seong Jin and Li, Jianlin and Daniel, Claus and Mohanty, Debasish and Nagpure, Shrikant and Wood, David L.},
  journal   = {Carbon},
  title     = {The state of understanding of the lithium-ion-battery graphite solid electrolyte interphase (SEI) and its relationship to formation cycling},
  year      = {2016},
  issn      = {0008-6223},
  month     = aug,
  pages     = {52--76},
  volume    = {105},
  doi       = {10.1016/j.carbon.2016.04.008},
  publisher = {Elsevier BV},
}

@Article{Peled2017,
  author    = {Peled, E. and Menkin, S.},
  journal   = {Journal of The Electrochemical Society},
  title     = {Review—SEI: Past, Present and Future},
  year      = {2017},
  issn      = {1945-7111},
  number    = {7},
  pages     = {A1703--A1719},
  volume    = {164},
  doi       = {10.1149/2.1441707jes},
  publisher = {The Electrochemical Society},
}

@Article{Huang2019,
  author    = {Huang, William and Attia, Peter M. and Wang, Hansen and Renfrew, Sara E. and Jin, Norman and Das, Supratim and Zhang, Zewen and Boyle, David T. and Li, Yuzhang and Bazant, Martin Z. and McCloskey, Bryan D. and Chueh, William C. and Cui, Yi},
  journal   = {Nano Letters},
  title     = {Evolution of the Solid–Electrolyte Interphase on Carbonaceous Anodes Visualized by Atomic-Resolution Cryogenic Electron Microscopy},
  year      = {2019},
  issn      = {1530-6992},
  month     = jul,
  number    = {8},
  pages     = {5140--5148},
  volume    = {19},
  doi       = {10.1021/acs.nanolett.9b01515},
  publisher = {American Chemical Society (ACS)},
}

@Article{Lehnert2025,
  author    = {Lehnert, Lukas and Lorenz, Martin and Juarez, Maria Fernanda and Schammer, Max and Nojabaee, Maryam and Schönhoff, Monika and Horstmann, Birger},
  journal   = {Journal of The Electrochemical Society},
  title     = {Combining Molecular Dynamics and Experimental Methods for the Parametrization of Binary Carbonate-Based Electrolytes},
  year      = {2025},
  issn      = {1945-7111},
  month     = may,
  number    = {5},
  pages     = {050523},
  volume    = {172},
  doi       = {10.1149/1945-7111/add381},
  publisher = {The Electrochemical Society},
}

@Article{Doyle1993,
  author    = {Doyle, Marc and Fuller, Thomas F. and Newman, John},
  journal   = {Journal of The Electrochemical Society},
  title     = {Modeling of Galvanostatic Charge and Discharge of the Lithium/Polymer/Insertion Cell},
  year      = {1993},
  issn      = {1945-7111},
  month     = jun,
  number    = {6},
  pages     = {1526--1533},
  volume    = {140},
  doi       = {10.1149/1.2221597},
  publisher = {The Electrochemical Society},
}

@Article{Fuller1994,
  author    = {Fuller, Thomas F. and Doyle, Marc and Newman, John},
  journal   = {Journal of The Electrochemical Society},
  title     = {Simulation and Optimization of the Dual Lithium Ion Insertion Cell},
  year      = {1994},
  issn      = {1945-7111},
  month     = jan,
  number    = {1},
  pages     = {1--10},
  volume    = {141},
  doi       = {10.1149/1.2054684},
  publisher = {The Electrochemical Society},
}

@Article{Latz2013,
  author    = {Latz, Arnulf and Zausch, Jochen},
  journal   = {Electrochimica Acta},
  title     = {Thermodynamic derivation of a Butler–Volmer model for intercalation in Li-ion batteries},
  year      = {2013},
  issn      = {0013-4686},
  month     = nov,
  pages     = {358--362},
  volume    = {110},
  doi       = {10.1016/j.electacta.2013.06.043},
  publisher = {Elsevier BV},
}

@Article{Gostick2019,
  author    = {Gostick, Jeff and Khan, Zohaib and Tranter, Thomas and Kok, Matthew and Agnaou, Mehrez and Sadeghi, Mohammadamin and Jervis, Rhodri},
  journal   = {Journal of Open Source Software},
  title     = {PoreSpy: A Python Toolkit for Quantitative Analysis of Porous Media Images},
  year      = {2019},
  issn      = {2475-9066},
  month     = may,
  number    = {37},
  pages     = {1296},
  volume    = {4},
  doi       = {10.21105/joss.01296},
  publisher = {The Open Journal},
}

@Article{Lehnert2024,
  author    = {Lehnert, Lukas and Nojabaee, Maryam and Latz, Arnulf and Horstmann, Birger},
  journal   = {ChemElectroChem},
  title     = {Challenges in Measuring Transport Parameters of Carbonate‐Based Electrolytes},
  year      = {2024},
  issn      = {2196-0216},
  month     = apr,
  number    = {11},
  volume    = {11},
  doi       = {10.1002/celc.202400056},
  publisher = {Wiley},
}

@Article{Petzl2014,
  author    = {Petzl, Mathias and Danzer, Michael A.},
  journal   = {Journal of Power Sources},
  title     = {Nondestructive detection, characterization, and quantification of lithium plating in commercial lithium-ion batteries},
  year      = {2014},
  issn      = {0378-7753},
  month     = may,
  pages     = {80--87},
  volume    = {254},
  doi       = {10.1016/j.jpowsour.2013.12.060},
  publisher = {Elsevier BV},
}

@Article{Talian2019,
  author    = {Sara Drvari{\v{c}} Talian and Jernej Bobnar and Anton Rafael Sinigoj and Iztok Humar and Miran Gaber{\v{s}}{\v{c}}ek},
  journal   = {The Journal of Physical Chemistry C},
  title     = {Transmission Line Model for Description of the Impedance Response of Li Electrodes with Dendritic Growth},
  year      = {2019},
  month     = {oct},
  number    = {46},
  pages     = {27997--28007},
  volume    = {123},
  doi       = {10.1021/acs.jpcc.9b05887},
  publisher = {American Chemical Society ({ACS})},
}

@Article{Bergstrom2021,
  author    = {Helen K. Bergstrom and Kara D. Fong and Bryan D. McCloskey},
  journal   = {Journal of The Electrochemical Society},
  title     = {Interfacial Effects on Transport Coefficient Measurements in Li-ion Battery Electrolytes},
  year      = {2021},
  month     = {jun},
  number    = {6},
  pages     = {060543},
  volume    = {168},
  doi       = {10.1149/1945-7111/ac0994},
  publisher = {The Electrochemical Society},
}

@Article{Traskunov2022,
  author    = {Traskunov, Igor and Latz, Arnulf},
  journal   = {Electrochimica Acta},
  title     = {Novel local fluctuation-preserving upscaling techniques for heterogeneous reaction-transport equation in porous media},
  year      = {2022},
  issn      = {0013-4686},
  month     = dec,
  pages     = {141248},
  volume    = {434},
  doi       = {10.1016/j.electacta.2022.141248},
  publisher = {Elsevier BV},
}

@Article{Traskunov2021,
  author    = {Traskunov, Igor and Latz, Arnulf},
  journal   = {Electrochimica Acta},
  title     = {Localized fluctuations of electrochemical properties in porous electrodes of lithium-ion batteries: Beyond porous electrode theory},
  year      = {2021},
  issn      = {0013-4686},
  month     = may,
  pages     = {138144},
  volume    = {379},
  doi       = {10.1016/j.electacta.2021.138144},
  publisher = {Elsevier BV},
}

@Article{Traskunov2021a,
  author    = {Traskunov, Igor and Latz, Arnulf},
  journal   = {Energy Technology},
  title     = {New Reduced‐Order Lithium‐Ion Battery Model to Account for the Local Fluctuations in the Porous Electrodes},
  year      = {2021},
  issn      = {2194-4296},
  month     = jan,
  number    = {6},
  volume    = {9},
  doi       = {10.1002/ente.202000861},
  publisher = {Wiley},
}

@Article{Campbell2019,
  author    = {Campbell, Ian D. and Marzook, Mohamed and Marinescu, Monica and Offer, Gregory J.},
  journal   = {Journal of The Electrochemical Society},
  title     = {How Observable Is Lithium Plating? Differential Voltage Analysis to Identify and Quantify Lithium Plating Following Fast Charging of Cold Lithium-Ion Batteries},
  year      = {2019},
  issn      = {1945-7111},
  number    = {4},
  pages     = {A725--A739},
  volume    = {166},
  doi       = {10.1149/2.0821904jes},
  publisher = {The Electrochemical Society},
}

@Article{Kolzenberg2022a,
  author    = {von Kolzenberg, Lars and Stadler, Jochen and Fath, Johannes and Ecker, Madeleine and Horstmann, Birger and Latz, Arnulf},
  journal   = {Journal of Power Sources},
  title     = {A four parameter model for the solid-electrolyte interphase to predict battery aging during operation},
  year      = {2022},
  issn      = {0378-7753},
  month     = aug,
  pages     = {231560},
  volume    = {539},
  doi       = {10.1016/j.jpowsour.2022.231560},
  publisher = {Elsevier BV},
}

@Misc{Philipp2026,
  author    = {Philipp, Micha C. J. and Köbbing, Lukas and Karger, Alexander and Jossen, Andreas and Latz, Arnulf and Horstmann, Birger},
  title     = {Physics-based modeling of cyclic and calendar aging of LIBs with Si-Gr composite anodes},
  year      = {2026},
  copyright = {Creative Commons Attribution 4.0 International},
  doi       = {10.48550/ARXIV.2604.26545},
  publisher = {arXiv},
}

\end{document}